\documentclass[]{pasj01}
\draft

\Received{}
\Accepted{}

\begin{document}

\title{
Solar jet-like features rooted in flare ribbons}

\author{Xiaohong \textsc{Li}\altaffilmark{1,2}
\thanks{ }}
\altaffiltext{1}{CAS Key Laboratory of Solar Activity, National
Astronomical Observatories, Chinese Academy of Sciences, Beijing
100101, China}
\altaffiltext{2}{School of Astronomy and Space Science, University
of Chinese Academy of Sciences, Beijing 100049, China}
\email{lixiaohong@nao.cas.cn}

\author{Jun \textsc{Zhang}\altaffilmark{1,2}}
\email{zjun@nao.cas.cn}

\author{Shuhong \textsc{Yang}\altaffilmark{1,2}}

\author{Yijun \textsc{Hou}\altaffilmark{1,2}}



\maketitle

\begin{abstract}
Employing the high spatio-temporal \emph{Interface Region Imaging
Spectrograph} 1330 {\AA} observations, we investigated the jet-like features
that occurred during the X8.2 class flare in NOAA active region (AR) 12673 on
2017 September 10. These jet-like features were rooted in the flare ribbons.
We examined 15 features, and the mean values of the lifetimes, projected widths,
lengths and velocities of these features were 87 s, 890 km, 2.7 Mm and 70 km s$^{-1}$,
respectively. We also observed many jet-like features which happened during
the X1.0 class flare on 2014 October 25. We studied the spectra at the base
of a jet-like feature during its development. The Fe XXI 1354.08 {\AA} line
in the corona displays blueshift, while the Si IV 1402.77 {\AA} line in the
transition region exhibits redshift, which indicates the chromospheric evaporation.
This is the first time that the jet-like features are reported to be rooted
in the flare ribbons, and we suggest that these jet-like features were driven
by the mechanism of chromospheric evaporation.
\end{abstract}

\section{Introduction}

Solar X-ray jets, extreme-ultraviolet (EUV) jets, surges are similar plasma
eruptions that are magnetically rooted in the photosphere and ejected into
the corona along open magnetic filed lines (e.g., \cite{Shibata1994,
Yoko1995, Rao2016}). Solar jets have been extensively
studied in many wavelengths, e.g., H$\alpha$ (\cite{Schm1995, Li2015}),
ultraviolet (UV; \cite{Chen2008}), EUV (\cite{Jiang2007, Innes2016}),
X-ray (\cite{Zhang2000}) and white light (\cite{Moore2011}).

Solar jets are ubiquitous transient features occurring in various solar environments,
such as coronal holes (\cite{Yang2011, Chen2012}),
active regions (ARs; \cite{Shimojo1996}) and quiet regions
(\cite{Hong2011, Pan2016}). Sometimes, jets are observed to be associated
with filament eruptions (\cite{Moore2010, Sterling2015}), coronal mass
ejections (CMEs, \cite{Shen2012}) and flares (\cite{Wang2012}).

The unceasing advances made on spatial and temporal resolution of data
obtained by different space missions have enriched our understanding of
different observational processes and numerical investigations behind solar
jets. However, the underlying driving mechanisms of solar jets have not been
understood completely. A generally accepted driving mechanism for jets is
magnetic reconnection between emerging magnetic fluxes and ambient magnetic
fields (e.g., \cite{Canf1996, Zhang2017}), which sometimes
is also accompanied by magnetic cancellation (\cite{Young2014}).
Furthermore, simulation works furnish evidence that coronal jets can be generated
by the mechanism of chromospheric evaporation (\cite{Shimojo2001,
Mi2003}).

There are also many small scale plasma eruptive phenomena such as
macrospicules and chromosphere dynamic structures like spicules and mottles.
These jet-like structures are thought to be different scale jets or
manifestations in different wavelengths of the jet phenomenon
(e.g., \cite{DeP2011, Tsir2012}). In this Letter,
exploring the X8.2 flare produced in AR 12673 and the X1.0 class flare in AR 12192,
we reported the jet-like features rooted in the flare ribbons for the first time.

\section{Observations and data analysis}

We adopted the observations from the Helioseismic and Magnetic Imager (HMI; \cite{Sche2012,
Schou2012}) and the Atmospheric Imaging Assembly (AIA; \cite{Lemen2012})
on board the \emph{Solar Dynamics Observatory} (\emph{SDO}; \cite{Pes2012}). The HMI full-disk
line-of-sight magnetograms with the pixel size of 0$\arcsec$.5 and the cadence
of 45 s were employed. AIA gives full-Sun images with high spatial resolution
(0.$\arcsec$6 pixel$^{-1}$, $\sim$ 430 km) and high temporal cadence (12/24 s) in seven
EUV wavelength bands and three UV-visible channels. On September 10, we used the
AIA 94 {\AA}, 131 {\AA}, 171 {\AA}, 193 {\AA}, and 304 {\AA} images, which have strong
responses to logarithmic temperatures (Kelvin) of about 6.8, 7.0 (and 5.6), 5.8, 6.2 and 4.7
respectively, from 15:30 UT to 16:30 UT to study the jets occurring during
the X8.2 flare.

The \emph{Interface Region Imaging Spectrograph} (\emph{IRIS}; \cite{DeP2014})
satellite provides simultaneous spectral and imaging observations of
the solar atmosphere. For the X8.2 class flare on 2017 September 10, we obtained
a series of \emph{IRIS} slit-jaw 1330 {\AA} images taken from 12:59:47 UT
to 19:23:38 UT with a pixel scale of 0$\arcsec$.333, a cadence of 9 s,
and a field of view (FOV) of 119$\arcsec$ $\times$ 119$\arcsec$. The 1330 {\AA}
channel contains emission from the strong C II 1334/1335 {\AA} lines formed
in the chromosphere and lower transition region as well as the continuum from the
photosphere and lower chromosphere. For the X1.0 class flare on 2014 October 25,
we obtained the \emph{IRIS} slit-jaw 1330 {\AA} images with a cadence of
16 seconds and a pixel scale of 0$\arcsec$.333. We also employed the
\emph{IRIS} spectral profiles of two windows at ``C II 1336" and
``Si IV 1394" with a cadence of 5 s.

\begin{figure}
\includegraphics[bb=42 258 540 613,clip,angle=0,width=\columnwidth]{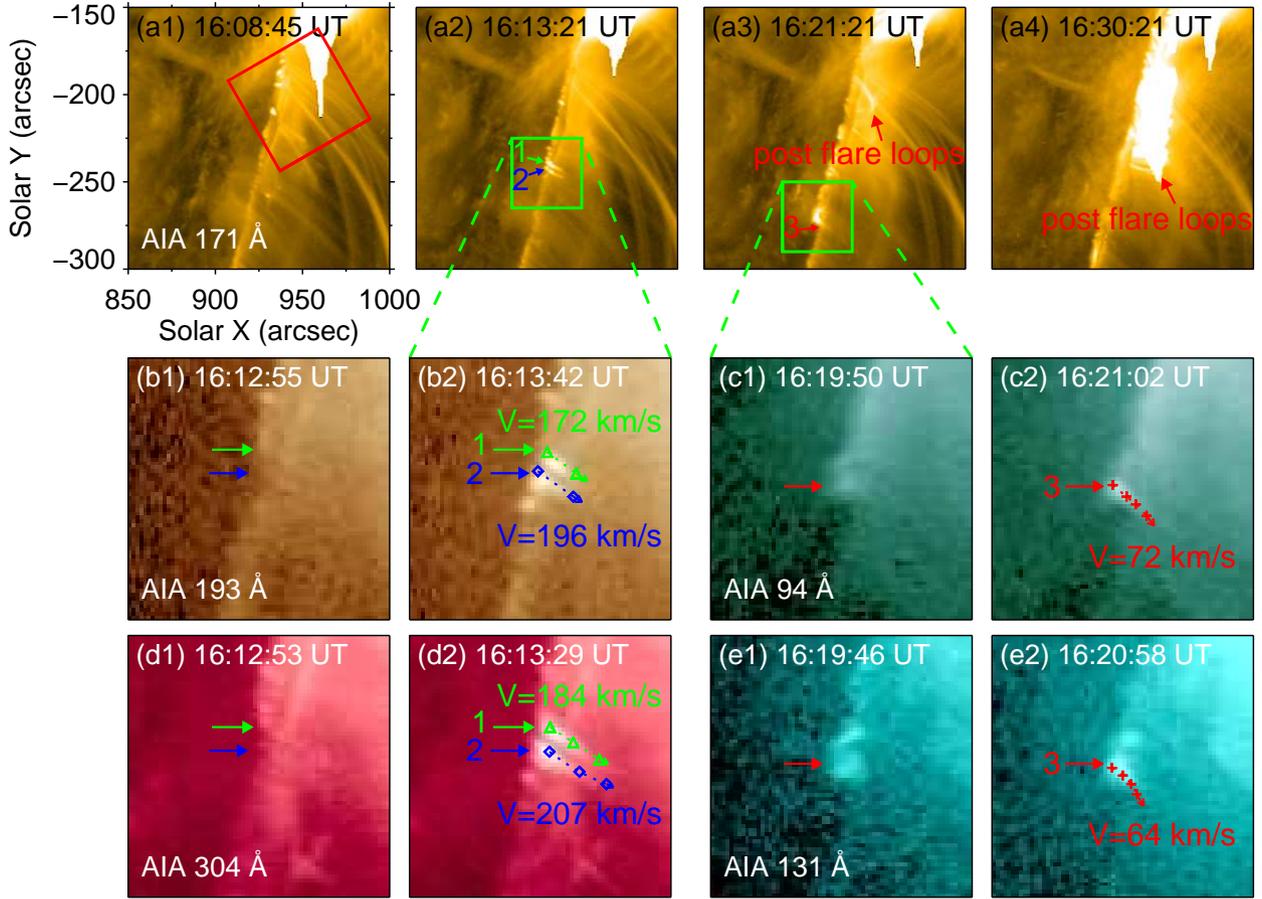}
\caption{Panels (a1)$-$(a4): AIA 171 {\AA} images showing the flare and
jet-like features on 10 September 2017. The red square in panel (a1) outlines
the FOV of Figure 2. Panels (b1) and (b2): AIA 193 {\AA} images displaying the
development of two jet-like features which is pointed out by the green and blue
arrows. Panels (d1) and (d2): evolution of same two features as shown in
panels (b1) and (b2) using AIA 304 {\AA} images. The green triangular symbols and
the blue diamond symbols indicate the trajectories of these two features.
Panels (c1)$-$(c2) and (e1)$-$(e2) are AIA 94 {\AA} images and 131 {\AA} images
displaying the development of a jet-like feature which is marked by the red
arrows, respectively. The red symbols display the trajectories of this feature.
An animation (Movie1) of this figure is available.
\label{fig1}}
\end{figure}

\section{Results and discussion}

\subsection{The X8.2 flare on 2017 September 10}

From 2017 September 4 to 10, AR 12673 produced 4 X-class, 27 M-class
and plenty of lower-class flares, becoming the most actively flaring
region of Cycle 24 (\cite{Yang2017}). On September 10, an X8.2 class
flare occurred in this AR. This X8.2 class flare started at 15:35 UT
and reached its peak at 16:06 UT. Using the AIA 171 {\AA} observations
(see Figures \ref{fig1} (a1)$-$(a4) and the animation of Figure \ref{fig1}),
we observed numerous jet-like features at the south side of this flare.
At 15:55 UT, these small scale jet-like features began to occur.
Along with the development of the flare, the jets appeared
farther south in succession since 16:09 UT, with an average
spreading speed of $\sim$ 45 km s$^{-1}$.
Two obvious jet-like features (labelled by ``1" and ``2" in panel (a2))
are displayed. These features emerged, rapidly grew and had
apparent velocities of 192 km s$^{-1}$ and 212 km s$^{-1}$
in 171 {\AA}, respectively. The apparent velocity here is the speed
of the propagating intensity front of the feature. The lengths of the two features were both
more than 9 Mm. The red arrows in panels (a3)$-$(a4) denote
the post flare loops in the 171 {\AA} observations. As the post flare
loops showed up, the jet-like features rooted in the footpoints
of these loops faded away (see panel (a4)). These jet-like features are also
distinct on the other AIA wavelength images. Panels (b1)$-$(b2) and panels (d1)$-$(d2)
are the AIA 193 {\AA} images and 304 {\AA} images displaying the development of
the features ``1" and ``2", individually. Panels (b1) and (d1) show the situation
prior to the feature onset. The green and blue arrows point to the places where the
features occurred. Panels (b2) and (d2) display the manifestations of these two features.
The trajectories of feature ``1" are indicated by the green triangular symbols.
The apparent velocity of feature ``1" was 172 km s$^{-1}$ in 193 {\AA}
and 184 km s$^{-1}$ in 304 {\AA}. The blue diamond symbols reveal the path of the feature ``2".
In 193 {\AA}, the velocity was 196 km s$^{-1}$, and in 304 {\AA}, feature ``2"
had a apparent velocity of 207 km s$^{-1}$. Panels (c1)$-$(c2) and panels (e1)$-$(e2) show
the development of the feature ``3" at 94 {\AA} and 131 {\AA} wavelengths, respectively.
The red arrows illustrate the position of feature ``3", the red symbols display the
trajectories of this feature, and the apparent velocities are 72 km s$^{-1}$ in
94 {\AA} and 64 km s$^{-1}$ in 131 {\AA}.

\begin{figure}
\includegraphics[bb=68 125 530 706,clip,angle=0,width=\columnwidth]{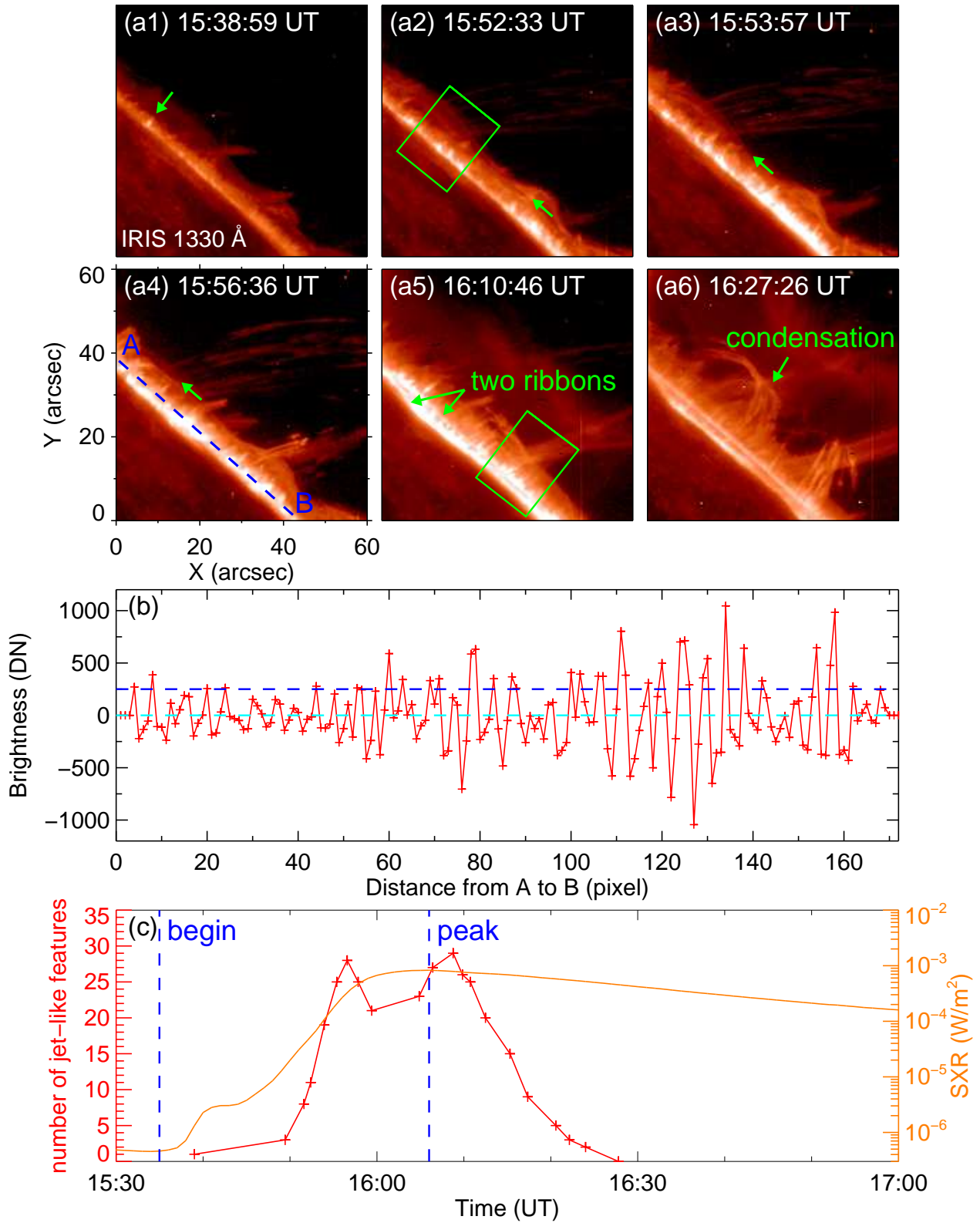}
\caption{Panels (a1)$-$(a6): \emph{IRIS} 1330 {\AA} images displaying the
development of the jet-like features rooted in the flare ribbons. The green arrow
in panel (a1) denotes the position where the brightening first occurred and the green
arrows in panels (a2)$-$(a4) indicate the propagating of the ribbons. The green
rectangle in panel (a2) outlines the FOV of Figures \ref{fig3}(a1)$-$(a4) and the green rectangle
in panel (a5) outlines the FOV of Figures \ref{fig3}(b1)$-$(b4). Panel (b): brightness
above the mean value across the slice ``A$-$B" in panel (a4). Panel (c): variations
of the number of the jet-like features during the flare presented by the red
cross symbols. The brown curve displays the variation of the \emph{GOES} soft X-ray 1$-$8 {\AA}
flux and the blue dotted lines indicate the begin and peak time of the flare.
An animation (Movie2) of this figure is available.
\label{fig2}}
\end{figure}

\emph{IRIS} also observed lots of jet-like features rooted in the flare ribbons
(see the animation of Figure \ref{fig2}) and the FOV of \emph{IRIS} is delineated
by the red square in Figure \ref{fig1}(a1). Figures \ref{fig2}(a1)$-$(a6) display
the development of these jet-like features observed in \emph{IRIS} 1330 {\AA}.
At 15:38 UT, small scale jet-like features (marked by the green arrow in panel (a1)) started
to occur. From 15:52 UT to 15:58 UT, the jet-like features on a flare ribbon brightened up
successively as indicated by the green arrows in panels (a2)$-$(a4). The second
ribbon can also be observed since 15:58 UT, and there were a lot of jet-like features
rooted in the both flare ribbons as shown in panel (a5). After 16:23 UT, there
was cold material which started to condense at the top of the loops and dropped to
the footpoints, identifying the structures of the post-flare loops. Since the condensation
showed up, these jet-like features located on the flare ribbons began to disappear (see panel (a6)).
To estimate the number of the jet-like features, we studied the brightness along the positions
identified by slice ``A$-$B" in panel (a4). The trend value is estimated using the
Single Gaussian fitting. Figure \ref{fig2}(b) displays the result of brightness minus
the trend value. The background value was about 1000 DN pixel$^{-1}$, and we
defined that a feature's brightness exceeded a quarter of the background brightness.
Therefore, there existed 28 jet-like features whose exceeded brightness were over 250 DN pixel$^{-1}$
(the blue line in panel (b)) on the 41 Mm length area. We estimate the number of the
jet-like features and its variations during the flare are presented in panel (c) with
the red cross symbols. The brown curve displays the variation of the \emph{GOES}
soft X-ray 1-8 {\AA} flux, and the begin and peak times of the flare
are indicated by the blue dotted lines. The jet-like features occurred after
the flare began, and the number of jets increased. After the peak of the flare,
the number of the jet-like features decreased and then disappeared.

\begin{figure}
\includegraphics[bb=50 250 512 563,clip,angle=0,width=\columnwidth]{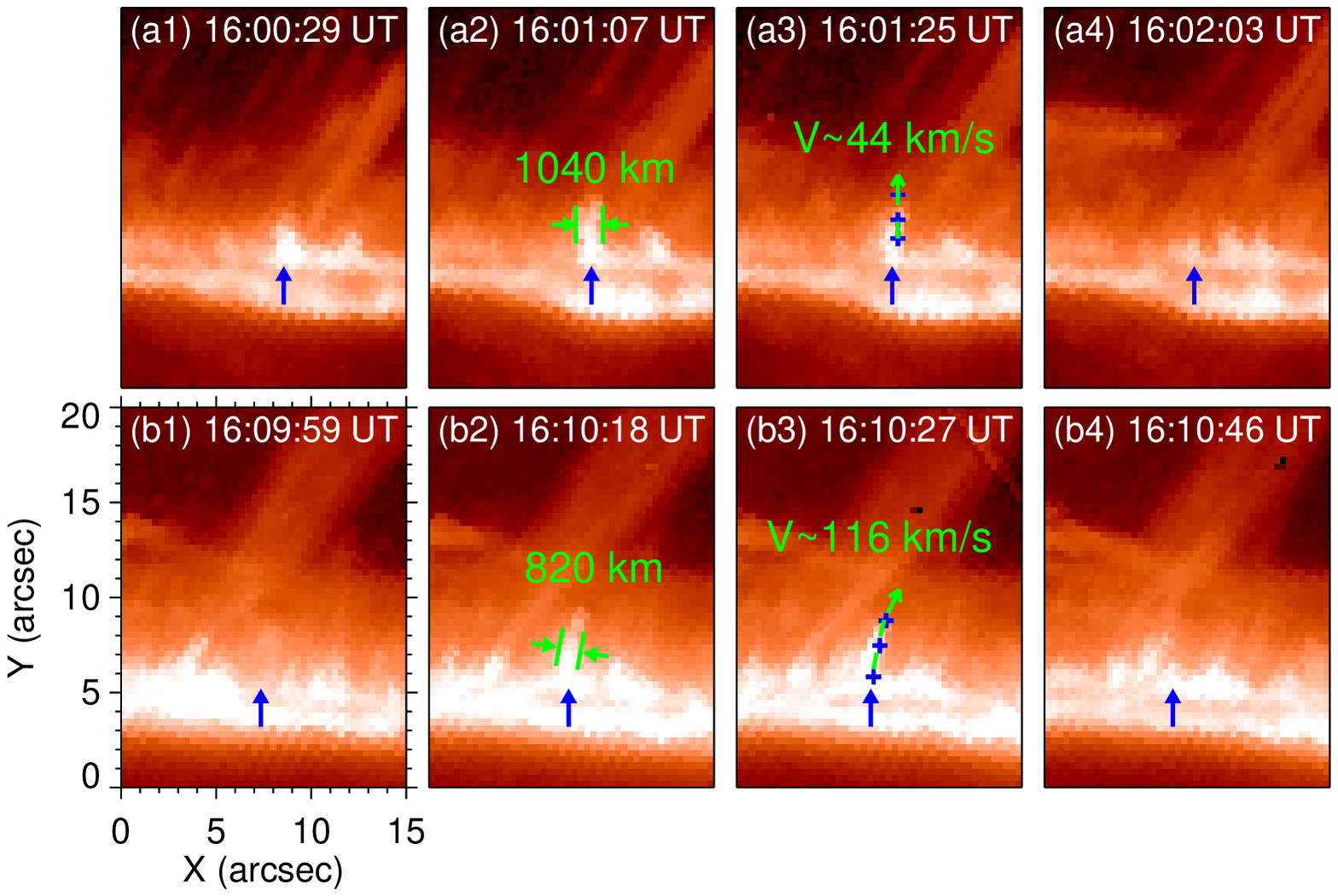}
\caption{\emph{IRIS} 1330 {\AA} images displaying the evolutions of two
jet-like features. The blue arrows denote the positions of the two features.
The blue crosses in panels (a3) and (b3) indicate the trajectories of the
bright points which we track to determine the velocities of these features.
\label{fig3}}
\end{figure}

Figure \ref{fig3} displays the evolution of two jet-like features observed in \emph{IRIS} 1330 {\AA} images.
The blue arrows in the upper/lower panels point out the positions of the first/second
feature we focus on. The first jet-like feature occurred at the area outlined
by the green rectangle in Figure \ref{fig2}(a2), which began at 15:59:52 UT, rose up from the
limb and then fell back. During its rising stage, the feature
had a apparent velocity of 44 km s$^{-1}$. The feature had a maximum width of 1040 km,
and lasted for about 140 s. At 16:09:59 UT, the second feature took place in
the area delineated by the green rectangle in Figure \ref{fig2}(a5). It ejected upward,
fell back, and then disappeared at 16:10:55 UT. The width of the second feature
was about 820 km, and the feature had a apparent velocity of 116 km s$^{-1}$.

We analyzed the properties of 15 isolated jet-like features. The lifetimes of
the jet-like features were 56 $\sim$ 184 s, with the mean value of 87 s. The
widths of the jet-like features were between 670 and 1340 km, and the lengths
of the jet-like features were 2.0 $\sim$ 3.8 Mm. All these features went upward
and had apparent velocities of 28 $\sim$ 116 km s$^{-1}$. Through
statistics, the mean values of width, length and velocity of the jet-like features
were 890 km, 2.7 Mm and 70 km s$^{-1}$, respectively. The \emph{IRIS} observed the
flare ribbon with a length of 51 Mm, and the occurrence of the jet-like features
lasted for about 20 minutes. Since there were 28 jet-like features on a 41 Mm length
flare ribbon and the mean lifetime of jet-like features was 87 s, we can estimate
that there arose approximately 480 jet-like features in the \emph{IRIS} observations.

The jet-like features rooted in the flare ribbons are reported for the
first time. The temporal and spatial relationships between the jet-like features
and flare suggest that these jet-like features may possess a different
driving mechanism, which is related to the onset of the two-ribbon flare.
According to the standard model of two-ribbon flares (\cite{Forbes2010,
Fle2011}), magnetic reconnection release energy, and this
energy accelerates non-thermal particles and heats the plasma near the
reconnection site. The combination of the thermal conduction and particles beams transport
the energy to the chromosphere through the relatively tenuous corona and stop in the
cooler, dense plasma. When the energy input to the chromosphere exceeds what can be shed
by radiative and conductive losses, chromospheric material is heated rapidly up to a
temperature on the order of 10 MK. The overpressure associated with the chromospheric
heating will drive the heated plasma upward to fill the flare loops (\cite{Priest2002}).
The evaporation upflows are often from the outer edges of the flare ribbons,
which may look like that the evaporation upflows are rooted in the flare ribbons.

This flare was located at the solar west limb, so we could directly observe
the vertical motions of the jet-like features against the solar surface.
These jet-like features were rooted in the flare ribbons and could be recognised
using the AIA EUV wavelengths. The velocities of jet-like features were 28 $\sim$
116 km s$^{-1}$ with the mean value of 70 km s$^{-1}$, consistent with previous
\emph{IRIS} spectroscopic observations in which the velocities of
evaporation upflows are mostly around 100 km s$^{-1}$ or smaller (e.g., \cite{Zhang2016,
Li2017}). Combining these observations with the theory, we suggest that these
jet-like features on the flare ribbon were driven by chromospheric evaporation.

\subsection{The X1.0 flare on 2014 October 25}

\begin{figure}
\includegraphics[bb=25 137 566 622,clip,angle=0,width=\columnwidth]{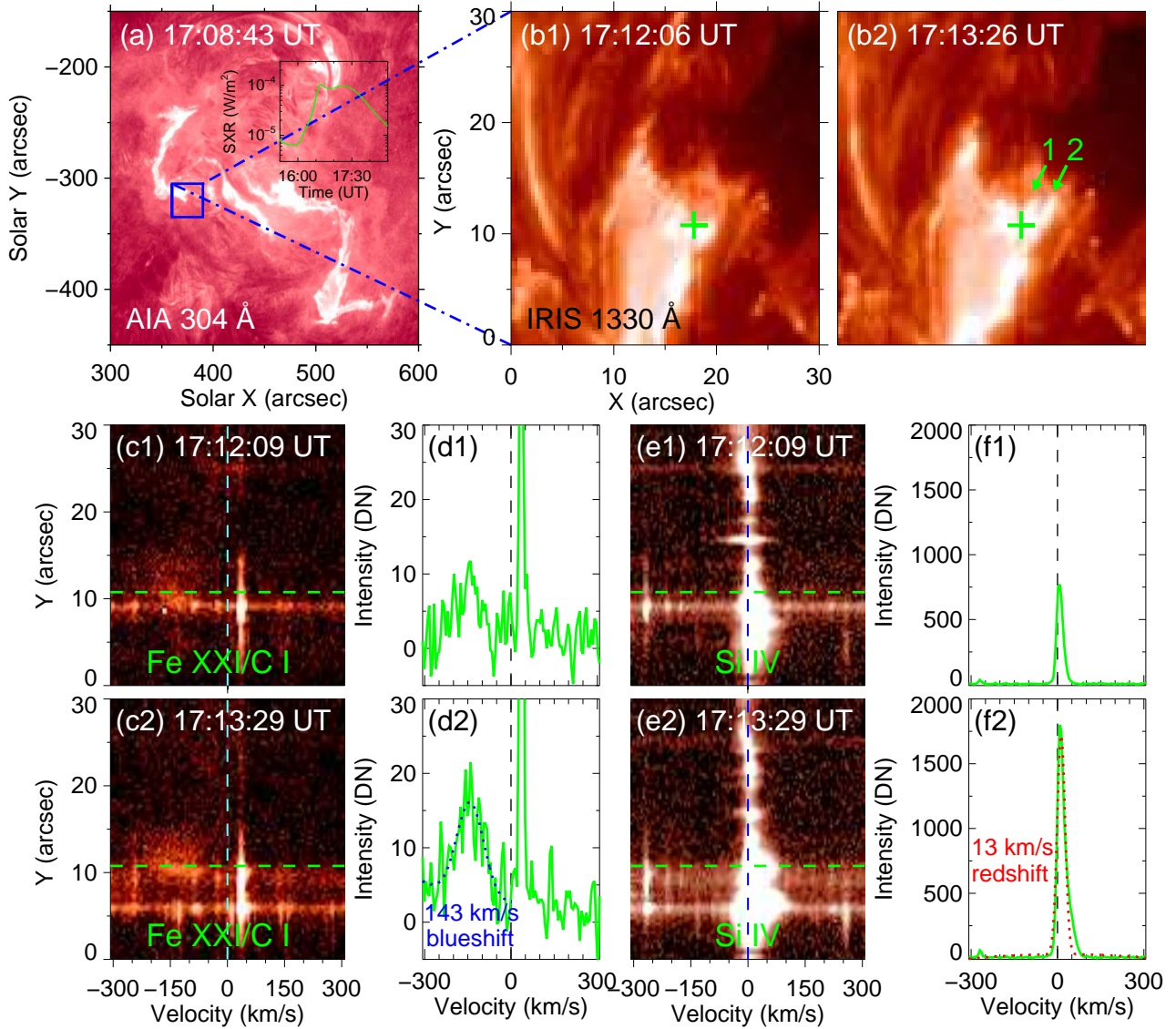}
\caption{(a): \emph{SDO}/AIA 304 {\AA} images displaying the X1.0 class flare
on October 25, 2014. (b1) and (b2): \emph{IRIS} 1330 {\AA} images displaying the evolution of
two jet-like features. The green symbols denote the positions where Doppler
shifts are measured. (c1) and (c2) are simultaneous appearances of the Fe XXI 1354.08 {\AA}
spectra in the slit range of (b1) and (b2) for Doppler velocities from
-310 km s$^{-1}$ to 310 km s$^{-1}$. (d1) and (d2) exhibit observed profiles
(green solid curves) at the selected locations in (c1) and (c2). The blue dashed
line in (d2) represents the Gaussian fitting to the blueshifted Fe XXI 1354.08 {\AA} feature.
(e1) and (e2) display simultaneous appearances of the Si IV 1402.77 {\AA}
spectra in the slit range of (b1) and (b2) for Doppler velocities from
-310 km s$^{-1}$ to 310 km s$^{-1}$. The green solid curves in (f1) and (f2)
are observed profiles at the selected locations in (e1) and (e2).
The red dashed line in (f2) exhibits the corresponding Gaussian fitting
to the Si IV 1402.77 {\AA} profile.
An animation (Movie3) of this figure is available.
\label{fig4}}
\end{figure}

Previous works on chromospheric evaporation mostly rely on spectroscopic
observations, since the best way to observe vertical flow in on-disk observations
is spectroscopy. The spectral signature of chromospheric evaporation
is the blueshift of hot emission lines, indicating the presence of hot and
fast (from tens of km s$^{-1}$ to several hundreds of km s$^{-1}$) plasma
upflows (\cite{Fisher1985, Tian2015}). We checked more \emph{IRIS} data
to look for the events in which similar jet-like features occurred and the
features were captured by the \emph{IRIS} slit. During the X1.0 class on 2014 October 25,
we found similar jet-like features. The X1.0 flare happened in AR 12192,
which produced 6 X-class and 29 M-class flares from October 18 to 29.
As shown in Figure \ref{fig4}(a), this flare is a two-ribbon flare. The GOES soft
X-ray 1$-$8 {\AA} flux showed that the X1.0 flare initiated at 16:55 UT
and reached its peak at 17:08 UT (see the green curve). The flare ribbons
could also be observed using the \emph{IRIS} 1330 {\AA} slit-jaw images.
During the rising stage of the flare, we observed several jet-like features
at one ribbon of this flare, and two of them are displayed in
Figures \ref{fig4}(b1)$-$(b2).

During the development of the jet-like feature ``1",
it passed through the \emph{IRIS} slit. At the base of the jet-like feature
(marked by the green cross symbols), we employed the \emph{IRIS} raster level 2 data,
which have been dark corrected, flat fielded, and geometrically corrected,
to measure the spectral information of this jet-like feature. The hot line
of Fe XXI 1354.08 {\AA} (log T $\sim$ 7.05) in the corona and the cool line
of Si IV 1402.77 {\AA} (log T $\sim$ 4.8) in the transition region were
chosen to investigate chromospheric evaporation (\cite{DeP2014,
Tian2014}). The nearby relatively strong neutral lines,
``O I" 1355.5977 {\AA} and ``S I" 1401.5136 {\AA} (\cite{Tian2015})
were used for absolute wavelength calibration, respectively.
As shown in Figures \ref{fig4}(c1)$-$(d2), there is significant enhancement at
the blue side of the Fe XXI 1354.08 {\AA} rest position.
There are several neutral and singly ionized lines, i.e.,
the C I line at 1354.29 {\AA}, the Fe II lines at 1353.02 {\AA},
1354.01 {\AA}, and 1354.75 {\AA}, the Si II lines at 1352.64 {\AA} and
1353.72 {\AA}, and some unidentified lines at 1353.32 {\AA} and
1353.39 {\AA}, are blended with the broad line of Fe XXI 1354.08 {\AA}.
However, these lines are narrow and cannot explain the bulk enhancement.
This emission feature is likely due to the blueshifted Fe XXI line.
The blueshift of the Fe XXI line became enhanced and more clear
as the jet-like feature developed ($\sim$ 143 km s$^{-1}$,
see panel (d2) and Movie3). As shown in panels (e1) and (e2),
the cool Si IV 1402.77 {\AA} line exhibits obvious redshift, and
the Doppler velocity relative to the background was about
13 km s$^{-1}$ (see panels (f1) and (f2)).
Our results agree with the spectroscopic observations that
explosive chromospheric evaporation is usually identified
by high speed blueshift ($\sim$ 100$-$400 km s$^{-1}$) in the
hot lines from corona, and accompanied by chromospheric condensation
which is distinguished by low speed redshift ($\sim$ 10$-$40 km s$^{-1}$)
in the cool lines from the upper chromosphere
and the transition region (\cite{Dos2013}).

\section{Conclusions}

We explored the X8.2 class flare in AR 12673 on 2017 September 10. Using the
\emph{SDO}/AIA 171 {\AA} observations, we discovered lots of jet-like features
occurring in succession along with the development of the flare ribbons.
The average spreading speed of these features was 45 km s$^{-1}$,
in accordance with the elongation velocity (ranging from 11 to
100 km s$^{-1}$) of the flare ribbons (\cite{Fle2004, Qiu2017}).
The footpoints of these jet-like features were located on the flare ribbons and
the apparent velocities of two most distinct jets were 153 km s$^{-1}$ and 192 km s$^{-1}$.
These jet-like features could also be distinguished in other EUV wavelengths.
Using the high tempo-spatial \emph{IRIS} 1330 {\AA} data, we also observed a
great deal of jet-like features rooted in the flare ribbons. During the development
of the flare, these jet-like features rose up from the limb and then fell back.
We studied 15 isolated features and their estimated lifetime, average projected width,
projected length and apparent velocity were 87 s, 885 km, 2.7 Mm and 70 km s$^{-1}$, respectively.
After 16:23 UT, cold material started to condense, dropping from the top to the footpoints
of the loops, and the jet-like features stopped occurring.

Using the \emph{IRIS} spectroscopic observations, we also investigated the jet-like features
appeared during the X1.0 flare in AR 12192 on 2014 October 25. We analysed the spectral
profiles of Fe XXI 1354.08 {\AA} and Si IV 1402.77 {\AA} at the base of a jet-like feature.
The hot Fe XXI line is blueshifted by $\sim$ 143 km s$^{-1}$, while the cool Si IV line
exhibits obvious redshift of $\sim$ 13 km s$^{-1}$, which is consistent with the scenario
of chromospheric evaporation.

Here, we directly observed the jet-like features with apparent motion
perpendicular to the flare ribbons, which is the direct observational
evidence of chromospheric evaporation.

\begin{ack}
We thank the referee for valuable suggestions.
This work is supported by the National Natural Science Foundations
of China (11533008, 11790304, 11673035, 11773039, 11673034, 11790300),
Key Programs of the Chinese Academy of Sciences (QYZDJ-SSW-SLH050),
and the Youth Innovation Promotion Association of CAS (2014043). The data
are used courtesy of \emph{IRIS}, \emph{SDO}, and \emph{GOES} science teams. \\
\end{ack}


\end{document}